\documentclass[aps,pre,onecolumn,epsf,graphicx,a4]{revtex4}
\usepackage[dvips]{graphicx}
\usepackage{amsfonts}
\usepackage{epsfig}
\usepackage{amsmath}
\usepackage{amssymb}

\def\pd2x{{\partial^2 \over \partial x^2}}

\usepackage{epsfig,latexsym}

\newcommand \bew {\begin{widetext}}
\newcommand \enw {\end{widetext}}

\begin{document}

\title{\bf\noindent 
The Centred Traveling Salesman at Finite Temperature}

\author{David Lancaster} 

\affiliation{
Harrow School of Computer Science, University of Westminster, 
Harrow, HA1 3TP, UK 
}
\date{8th September 2006}
\begin{abstract}
A recently formulated statistical mechanics method is used
to study the phase transition occurring in a generalisation
of the Traveling Salesman Problem (TSP) known as the centred TSP.
The method shows that the problem has clear signs of a crossover,
but is only able to access (unscaled) finite temperatures
above the transition point. The solution of the problem
using this method displays a curious duality.
\end{abstract}

\maketitle
\vspace{.2cm} \pagenumbering{arabic}

\section{Introduction}

In a recent paper, Lipowski and Lipowska~\cite{lipo2}
considered a generalisation of the geometric TSP by adding a new
term to the cost function:
\begin{eqnarray}
H &=& (1-\alpha) H_L + \alpha H_C\\
H_L &=& \sum_i V_L({\bf r}_{i+1},{\bf r}_i)=\sum_i |{\bf r}_{i+1} - {\bf r}_i|\\
H_C &=& \sum_i V_C({\bf r}_{i+1},{\bf r}_i)=\sum_i |{\bf r}_{i+1} + {\bf r}_i|
\end{eqnarray}
The familiar term $H_L$ of the traditional TSP 
is the total path length of a cyclic tour defined by some 
permutation of the $N$ city indices.
The new term $H_C$ is a measure of the distance of the 
tour from the centre of the geometry. Here we shall only
consider the two-dimensional problem in a square domain,
and shall take this centre, the origin of the city
coordinates, ${\bf r}_i$, to be the centre of the square.
The parameter $\alpha$, $0\le \alpha\le 1$, specifies the
relative strength of each term.

The authors of \cite{lipo2} gave example motivations for
the centred model, but the fact that they found a transition
as $\alpha$ is varied and were able to identify it as
a complexity transition provides sufficient interest. 
Specifically, they used simulated annealing techniques to numerically
study zero temperature optimum configurations. While
these configurations appear similar to standard TSP tours for small $\alpha$,
the picture is rather different at large $\alpha$.
For $\alpha = 1$, rather than connecting near neighbour cities 
(${\bf r}_{i+1} \sim {\bf r}_i$), the
tour connects cities lying on points that are almost inverted
with respect to the origin (${\bf r}_{i+1} \sim -{\bf r}_i$).
This change in character can be seen directly though an observable
measuring the angle between successive links of the tour:
at small $\alpha$ the angle is not especially biased
in any direction (though boundary effects cause a slight
preference for acute angles), while at large $\alpha$
there is a distinct preference for links to double back 
on themselves.
Sharp changes in quantities such as $\langle H_L\rangle$ 
as $\alpha$ is varied suggest a phase transition.
Moreover, this transition was identified as a
complexity transition~\cite{SAT}, since the simulated annealing schedule
required to reach optima in the (large $\alpha$) C-phase
is much less stringent than that required in the 
(small $\alpha$) L-phase, indicating a difference in
characteristic difficulty of the problem as the
transition is crossed.
Complexity transitions have recently been of strong interest 
to computer scientists, combinatorial optimists and physicists,
and although a transition has already been identified \cite{Korp} in
other versions of the TSP,
the geometric basis and simplicity of this centred model are very appealing.

Many questions about the nature of the transition remain, and
the aim of this paper is to investigate the model using analytic
techniques recently developed to study the statistical mechanics
of TSP-like models~\cite{us}. These techniques allow a full
solution of the theory in the region of high city density and
finite temperature.
In particular, expectation values for
the path length $\langle H_L\rangle$ and its fluctuations
can be computed and these
provide  evidence for a smoothed transition
at finite temperature.
The technique is not able to access the optima at low temperature
and we cannot use it to compute critical indices for example. 
Nonetheless, our methods produce other insights such as
a duality; $\alpha \leftrightarrow (1-\alpha)$ between thermodynamic
quantities that may have a role beyond the region of applicability of
the technique.

\section{Formalism}

The analytic techniques we use to solve the model were 
derived in \cite{us}  using a functional formalism
and have since been placed on a firm basis via a discrete approach
 \cite{discrete}.
The continuum equations for a system of unit area and
a flat distribution of cities allow the 
free energy to be written in terms of a quantity $s({\bf r})$
which is defined by an integral equation,
\begin{eqnarray}
\beta F &=&  -2N \int d^2r\,  \log s({\bf r}) - N \log N \\
s({\bf r}) &=& \int d^2r\, {1\over s({\bf r}')}
\exp\left(-\beta  \left[(1-\alpha)V_L({\bf r},{\bf r}')+\alpha V_C({\bf r},{\bf r}') \right]\right).
\label{eq:continuum}
\end{eqnarray}
Observables can now be obtained by standard thermodynamic relations.
For example the expectation values for the total length
of the path and its fluctuations are given by,
\begin{eqnarray}
\langle H_L \rangle &=& 
N \int d^2r d^2r'\, {V_L({\bf r},{\bf r}')  \over s({\bf r})s({\bf r}')}
\exp\left(-\beta  \left[(1-\alpha)V_L({\bf r},{\bf r}')+\alpha V_C({\bf r},{\bf r}') \right]\right)\\
\langle (H_L-\langle H_L \rangle )^2 \rangle &=& 
{\alpha\over\beta}{\partial \langle H_L \rangle \over \partial \alpha}
-{\partial \langle H_L \rangle \over \partial \beta}
\label{eq:expectations}
\end{eqnarray}
These expressions are proportional to $N$ in contrast to the
expected dependence of the optimum configurations. 
For example, the optimum path length for the
standard ($\alpha=0$) TSP grows as $\sqrt{N}$, and for $\alpha > 0$
has  more complex $N$ dependence reported in~\cite{lipo2}.
Along with the non-extensive $N\log N$ entropy, this scaling makes it
already clear that the region within which the technique 
is valid is restricted.
Similar formulae hold for expectations of $H_C$, but in all cases 
we find that the solution of the integral equation (\ref{eq:continuum}), 
obeys a duality relation:
\begin{equation}
s_{1-\alpha}({\bf r}) = s_{\alpha}({\bf r})  = s_{\alpha}(-{\bf r}) 
\end{equation}
This duality is a consequence of the symmetry that 
inverts with respect to the centre,
${\bf r}\to -{\bf r}$, and interchanges $H_L\leftrightarrow H_C$.
As a result, $F(1-\alpha) = F(\alpha)$ so expectations
of $H_C$ are simply related to those of $H_L$, for
example, $\langle H_C(\alpha)\rangle
= \langle H_L(1-\alpha)\rangle$. 
One immediate result of the duality is that 
if any transition occurs, it must be at $\alpha_c = 1/2$.
We shall return to discuss
whether this duality holds outside the range of validity
of the equations (\ref{eq:continuum}).

\begin{figure}
\epsfxsize=0.5\hsize \epsfbox{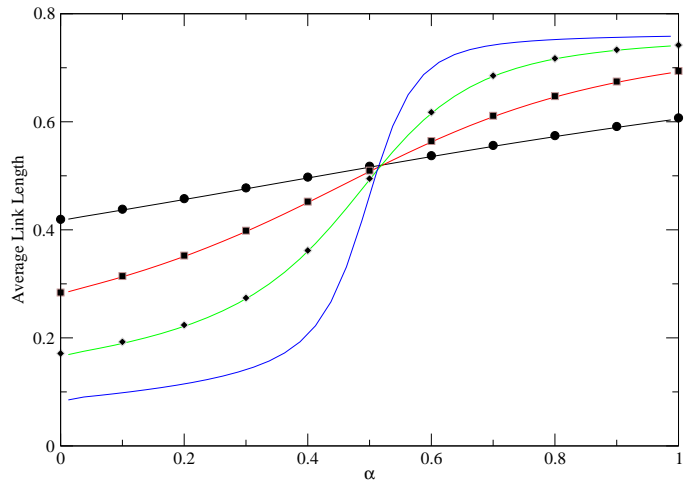}
\caption{The average link length $(1/N)\langle H_L \rangle$ for 
the centred TSP according to the formalism of \cite{us}
shown for $\beta = 2.0, 5.0, 10.0, 20.0$ with increasingly sharp step.
Points (omitted at $\beta=20.0$) are from Monte Carlo simulations.
}
\label{Lvev}
\end{figure}


It has not been possible  to analytically solve the 
integral equation (\ref{eq:continuum}), but an iterative procedure
on a discretised version converges rapidly and is numerically stable.
Monte Carlo simulations independently confirm the results
and are shown in the figures below. 
At the lowest temperatures it becomes hard to equilibrate the
Monte Carlo and we omit the $\beta=20.0$ data.
In figure (\ref{Lvev}) the expectation value of the length per link
is shown as a function of $\alpha$, for a variety of temperatures.
Evidently this tends towards a step function and is a smoothed out 
version of the zero temperature result shown in reference \cite{lipo2}.

\begin{figure}
\epsfxsize=0.5\hsize \epsfbox{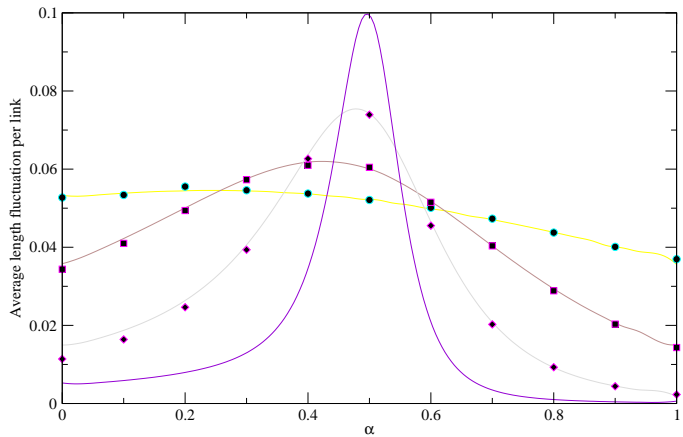}
\caption{The average per link length fluctuations
$(1/N)\langle (H_L-\langle H_L \rangle )^2 \rangle$ for
the centred TSP according to the formalism of \cite{us}
shown for $\beta = 2.0, 5.0, 10.0, 20.0$ with increasingly pronounced peak.
Points (omitted at $\beta=20.0$) are from Monte Carlo simulations.
}
\label{Lfluct}
\end{figure} 

Fluctuations in the path length also grow as $N$ and figure
(\ref{Lfluct}) shows how
$(1/N)\langle (H_L-\langle H_L \rangle )^2 \rangle$
changes with $\alpha$. This quantity is numerically more delicate
than the simple average, but although it can be computed by
solving linear equations for 
$\partial s/\partial \beta$ and $\partial s/\partial \alpha$,
we have simply evaluated the expression in 
(\ref{eq:expectations}) with discrete derivatives.
The Monte Carlo also suffers from large sample to sample fluctuations 
at low temperature.
As the temperature is reduced, a peak starts to appear at $\alpha_c$.

Other quantities such as correlations between
angles between successive links have also been computed
within this formalism and agree with Monte Carlo simulations. 
The computations are interesting in that they show how to
generalise the methods of \cite{us} to compute correlations
along the path, however they do not throw any further light
on this problem and its transition.

\section{Low Temperature Limit}

Although no general analytic solution of the integral equations
has been found, the low temperature limit can be understood using
a saddle point at large $\beta$. Because the 
formalism does not scale temperature~\cite{tspps1}
this low temperature limit does not correspond to the TSP optimum. 

Writing $s_\alpha({\bf r}) = t_\alpha({\bf r}) e^{-\beta w_\alpha({\bf r})}$,
the stationarity condition of the exponent at large $\beta$ is:
\begin{equation}
w_\alpha({\bf r}) = min_{{\bf r}'}\left[
(1-\alpha) |{\bf r} - {\bf r}'|
+\alpha |{\bf r} + {\bf r}'|
-w_\alpha({\bf r}')\right]
\end{equation}
A solution exists with $w_\alpha({\bf r})$ linear in ${\bf r}$,
matching the tendency of the iterative numerical solutions
at low temperature.
The prefactor $t_\alpha({\bf r})$, can also be determined
based on linear rather than the usual quadratic fluctuations
\cite{us}.
The resulting solution obeys duality:
\begin{equation}
s_\alpha({\bf r}) =
\left\{ 
\begin{array}{ll}
{\sqrt{2\pi}\over \beta (1-\alpha)} 
\exp \left( -\beta\alpha r\right) 
& 0 < \alpha \le 1/2 \\
{\sqrt{2\pi}\over \beta \alpha} 
\exp \left( -\beta(1-\alpha) r\right) 
& 1/2 < \alpha \le 1 
\end{array}
\right.
\end{equation}
The resulting expectation values have sharp changes at the
transition point. For example,
\begin{equation}
{1\over N}\langle H_L \rangle =
\left\{ 
\begin{array}{ll}
{2\over \beta (1-\alpha)} 
\exp \left( -\beta\alpha r\right) 
& 0 < \alpha \le 1/2 \\
2 \int d^2 {\bf r}\, |{\bf r}|
= \left(\sqrt{2}+\log(1+\sqrt{2})\right)/3
& 1/2 < \alpha \le 1 
\end{array}
\right.
\end{equation}
The length fluctuations develop a delta function singularity
at $\alpha_c$, with leading coefficient proportional to
$1/\beta$,
\begin{equation}
{1\over  N}\langle (H_L-\langle H_L \rangle )^2 \rangle
=
{1\over\beta}\left({\left(\sqrt{2}+\log(1+\sqrt{2})\right)\over 6} 
- {2\over\beta}\right)
\delta(\alpha-\alpha_c)
+{2\over \beta^2 (1-\alpha)^2}\theta (\alpha_c-\alpha)
\end{equation}
The height of the fluctuation peak as determined by
the iterative solution of the integral equation
at fixed $\alpha_C$
appears first to grow as $\beta$ as the temperature is reduced, 
but 
eventually this $1/\beta$ behaviour is observed.

\section{Conclusions}

The method of \cite{us} for studying the statistical mechanics of
TSP-like problems has been applied to the centred TSP. 
Both numerical solutions of the equations at finite temperature and the
limiting behaviour at large $\beta$ provide evidence
supporting the transition observed though numerical study of 
optimum configurations in \cite{lipo2}. The method
is not able to reach the region where optimal configurations
dominate and the character of the evidence
is a smoothed signal of the transition indicating that the
transition point lies at a temperature below that accessible
by this technique.

A noteworthy feature of the solutions is a duality, 
$F(1-\alpha)=F(\alpha)$, relating observables
on either side of the transition. This duality can be
understood in terms of the density of states: 
the number of cyclic tours with $H_L$ between $E$ and
$E+dE$ is the same as the number of tours with
$H_C$ between $E$ and $E+dE$,
where in this regime of high city density and high temperature,
$E$ should not be too small.
This kind of relationship also exists between
a particular pair of one-dimensional Hamiltonians~\cite{twoHs}
in which the configurations are also 
specified by permutations of $N$ indices. 
However, in that case, even at finite $N$, 
a map between a configuration with a certain value
of one Hamiltonian and another configuration with the same
value for the other Hamiltonian was demonstrated.
A brief consideration of particular instances of
the $N=4$ case shows that this precise relationship  cannot hold 
for the centred two-dimensional TSP, so the duality must
only be valid in the large $N$ limit.
Within the large $N$ limit, 
even if the duality  holds for optima and near optima
outside the regime of high temperature,
the lack of a clear map prevents one from using
the $\alpha=1$ model on the easy side of the transition
to solve the hard TSP at $\alpha = 0$.

\vskip 0,5 truecm

\noindent{\bf Acknowledgement:} I would like to
acknowledge a Discipline Hopping Award from the EPSRC.
 
\pagestyle{plain}


\baselineskip =18pt
\begin{thebibliography}{0}
\bibitem{lipo2}{A.~Lipowski and D.~Lipowska, 
Phys. Rev.{\bf E} 71, 067701 (2005)}
\bibitem{SAT}{
T.~Hogg, B.A.~Huberman and C.~Williams, Artif. Intell. {\bf 81}, 1 (1996);
O.C. Martin, R. Monasson and R. Zecchina, Theoretical 
Computer Science {\bf 265}, 3 (2001).
}
\bibitem{Korp}{
W.~Zhang and R.E.~Korf,
Artificial Intelligence {\bf 81}, 223 (1996);
P.~Cheeseman, B.~Kanefsky and W.M.~Taylor,
Proc AAAI-93, Washington, 21 (1993)}
\bibitem{us}{D.S.~Dean, D.~Lancaster and S.N.~Majumdar, 
J. Stat. Mech. L01001 (2005);    
{\it ibid}. Phys. Rev. E {\bf 72} 026125 (2005).}
\bibitem{discrete}{D.S.~Dean and D.~Lancaster, in preparation.}
\bibitem{tspps1}{J. Vannimenus and M. M\'ezard, J. Physique Lett.
{\bf 45} L1145 (1984).}
\bibitem{twoHs}{D.~Lancaster, J. Phys. A. {\bf 37}, 1125 (2004).}
\end{thebibliography}
\end{document}